\documentclass[]{tPHL2e}
\usepackage{graphicx}

\newcommand{\alt}{ \stackrel{<}{_{\sim}} }
\newcommand{\prl}{{\it Phys.\ Rev.\ Lett.} }

\begin{document}
\markboth{O.M. Braun and E. Tosatti}{Rack-and-pinion effects in molecular rolling friction}
\title{Rack-and-pinion effects in molecular rolling friction}
\author{Oleg M. Braun$^{\rm a,b}$$^{\ast}$\thanks{
$^\ast$Corresponding author. Email: obraun@iop.kiev.ua
\vspace{6pt}}
and
Erio Tosatti$^{\rm b,c,d}$\\\vspace{6pt}
$^{\rm a}${\em{Institute of Physics, National Academy of Sciences of Ukraine, 03028 Kiev, Ukraine}};
$^{\rm b}${\em{International School for Advanced Studies (SISSA), Via Beirut 2-4, I-34014 Trieste, Italy}};
$^{\rm c}${\em{CNR-INFM Democritos National Simulation Center, Via Beirut 2-4, I-34014 Trieste, Italy}};
$^{\rm d}${\em{International Centre for Theoretical Physics (ICTP), P.O. Box 586, I-34014 Trieste, Italy}}
\\\vspace{6pt}\received{...} }

\maketitle
\begin{abstract}
Rolling lubrication with spherical molecules working as ``nanobearings''
has failed experimentally so far, without a full understanding
of the physics involved and of the reasons why. Past model simulations
and common sense have shown that molecules can only roll when they are not too
closely packed to jam. The same type of model simulations now shows in addition
that molecular rolling friction can develop deep minima once the molecule's
peripheral ``pitch'' can match the substrate periodicity, much as ordinary
cogwheels do in a rack-and-pinion system. When the pinion-rack
matching is bad, the driven molecular rolling becomes discontinuous and noisy,
whence energy is dissipated and friction is large. This suggests
experiments to be conducted by varying the rack-and-pinion matching. That
could be pursued not only by changing molecules and substrates, but also
by applying different sliding directions within the same system, or by
applying pressure, to change the effective matching.


\end{abstract}

\section{Introduction}
\label{intro}

The problem of designing nanomechanical devices, in particular, of reducing friction
by means of nano- and micro-bearings~\cite{F1960,D1992} is nowadays a hot one.
The idea of using spherical molecules such as fullerene C$_{60}$
as molecular nanobearings gave rise to experimental attempts~\cite{78910, LTXLX2003, CK2006}
as well as Molecular Dynamics (MD)
simulations~\cite{LGG2004,SITM2007,KH2004,KHFBHK2005}.
Other potentially rolling molecules
were considered as additives in oil lubricants, and predicted to provide
interesting tribological properties~\cite{GHET2004}.
The simulations indicated that ball-shaped molecules may either slide or rotate
over a surface, depending on the substrate and the position of the molecule, and
suggested that the rolling configuration should indeed attain extremely
low friction~\cite{LGG2004,SITM2007}, with a predicted friction coefficient of the order
$\mu \sim 0.01-0.02$~\cite{KH2004} or smaller~\cite{SITM2007}.
All experimental attempts however met with scarce success so far.
A single C$_{60}$ molecule confined between solid substrates
should begin to roll under a torque of order
$10^{-19}$~Nm~\cite{MKS2003}. However, C$_{60}$ molecules
generally condense to form close-packed islands,
their two dimensional (2D) crystalline structure exhibiting 
order at low temperatures. Through a first-order orientational
order-disorder transition~\cite{141516} at $T = T_m \approx 260$ K, the molecules
actually unlock to attain seemingly free rotation at $T > T_m$.
At $T_m$ there is indeed an abrupt change in friction~\cite{LTXLX2003}, but
the lowest friction coefficient is still of order
$\mu \sim 0.15$~\cite{78910,LTXLX2003}, worse than with traditional oil-based lubricants.
Coffey and Krim~\cite{CK2006} undertook a quartz crystal microbalance study
of one or two C$_{60}$ monolayers adsorbed on Ag(111) or Cu(111).
There are no rotations in a C$_{60}$ layer on Cu(111), and only a slow change
of molecular orientations for a C$_{60}$ layer on Ag(111). In a bilayer,
C$_{60}$ molecules in the top layer rotate freely at 300 K, as expected.
Nonetheless a molecularly thin methanol film deposited over the C$_{60}$
bilayer failed to show either the expected low friction, or any essential difference
between these systems. Altogether, these results apparently doom the idea
to replicate the bearing concept at the nanoscale. The questions why, and how
this could be overcome are still quite open.

First of all, even a small amount of charge transfer and/or bonding
between C$_{60}$ and the metal substrate may hinder the rolling; therefore, chemically
inactive, insulating substrates should be preferred to metals. A second
handicap lies in the close packing of molecules. Balls in bearings are arranged
so as to prevent contact, but rolling molecules in a monolayer are in
close contact, hindering, even in the apparently freely rotating phase,
their mutual rolling and jamming the same way mutually ingrained rolling
cogwheels would. As was found in earlier simulations~\cite{B2005},
a way to avoid jamming is to lower dramatically the molecular coverage
well below a monolayer -- indicatively to coverages $\alt 0.3$ -- to approach
the limiting rolling friction of a single molecule. A third element, not
investigated so far, is a possible role of molecule-substrate matching.
Using a single molecule model as our starting point, we now find that
this element is unexpectedly important.

In our ordinary, macroscopic world, the main source of rolling
friction is deformation, since both substrate and roller
elastically or plastically deform at the contact. The deformation energy
is partly released and lost in the form of bulk frictional heat when the
roller moves on~\cite{P0}. By designing the bulk so that dissipation is poor in the right
frequency range, rolling friction can be made $10^2$ to $10^3$ times lower
than sliding friction, since the latter implies breaking and re-forming
of slider-substrate bonds. As the roller size is decreased however, adhesion
grows in importance, eventually becoming the main source of friction.
To rotate a molecule, one has to break the molecule-substrate bonds
one side of the molecule and create new bonds on the opposite side;
hence, molecular rolling friction might not be much lower than sliding friction.

Our goal is to understand what physics may yield the lowest friction coefficient
attainable for molecular rolling and which system parameters might provide it.
In the end, we find that nanobearings might indeed work as well as macroscopic
ones, but one has to choose properly the macroscopic counterpart, which here
turns out to be a perfect rack-and-pinion matching as in cogwheels. Because we are
interested in general trends, we presently explore a minimal 2D model, which
allows us to span a large number of parameters, and also provides an easier
visualization of all processes.

\section{Model}
\label{model}

Consider two substrates with lubricant molecules in between, all elements made
up of classical point particles (atoms). Atoms can generally move in the $(x,y)$ plane,
where $x$ is the sliding direction and $y$ is perpendicular to the substrates.
The substrates, pressed together by a load force $F_l = N_s f_l$,
consist of rigid atomic chains of length $N_s$ and lattice constant
$R_s$, so that the system size in the sliding direction is $L_x = N_s R_s$ and
the total mass of the substrate is $N_s m_s$ (we use periodic boundary condition
along $x$). The bottom rigid substrate is fixed at $x=y=0$,
the top one is free to move in both $x$ and $y$, while driven along $x$
through a spring of elastic constant $k_s$ moving with speed $v_s$.
The spring force $F$, whose maximum value before the onset of sliding measures the static
friction force $F_s$, and whose average during smooth motion $F_k = \langle F \rangle$
is the kinetic friction force, is monitored
during simulation (throughout the paper we will normalize forces per
substrate atom $f=F/N_s$). Thus, our model is a 2D variant of a typical
experimental setup in tribology \cite{P0,BN2006}.
Between the substrates we place circular (``spherical'') lubricant molecules.
As in Ref.~\cite{B2005}, each molecule consists of one central atom
and of $L$ atoms on circle of radius $R_m = R_{ll} /2 \sin(\pi/L)$
so that their chord distance is $R_{ll}$. All atoms interact via a
12-6 Lennard-Jones (LJ) potential
$V_{\rm LJ} (r)=V_{\alpha \alpha^{\prime}} \left[ \left( {R_{\alpha \alpha^{\prime}}}/{r}
\right)^{12} -2 \left( {R_{\alpha \alpha^{\prime}}}/{r} \right)^{6} \right]$,
where $\alpha,\alpha^{\prime} = s$ or~$l$ for the substrate or molecular
atoms respectively. The latter are additionally coupled to the central atom
by stiff springs of the elastic constant $K_m$,
$V_{\rm stab} (r)={1\over2} K_m (r-R_{\rm stab})^2$,
where the distance $R_{\rm stab} = R_m + (12 \, V_{ll} /K_m R_m)
\left[ (R_{ll} /R_m)^{6} - (R_{ll} /R_m)^{12} \right]$
is chosen so that the total potential
$V_{\rm LJ} (r) + V_{\rm stab} (r)$ is minimum at $r=R_m$.
With $K_m =100$ the resulting stiff molecular shape
resisted destruction during the simulations.
Thus, the lubricant-lubricant interaction is described by the parameters
$V_{ll}$ and $R_{ll}$, while the lubricant-substrate interaction, by
$V_{sl}$ and $R_{sl}$
(direct interaction between the top and bottom substrates is omitted).
We use dimensionless units, where $m_s = m_l =1$,
$R_{ll} =1$, and the energy parameters $V_{\alpha \alpha^{\prime}}$
takes values around $V_{\alpha \alpha^{\prime}} \sim 1$.
Because a 2D model cannot reproduce even qualitatively the phonon spectrum of a 3D system,
and because frictional kinetics is generally diffusional rather than inertial,
we use Langevin equations of motion with Gaussian random forces
corresponding to temperature $T$, and a damping force
$f_{\eta, x} = -m \, \eta (y) \, \dot{x} -m \, \eta (Y-y) \, (\dot{x} - \dot{X})$,
where $x,y$ are the atomic coordinates and
$X,Y$ are the coordinates of the top substrate
(the force $f_{\eta, y}$ is defined in the same way).
The viscous damping coefficient is assumed to decrease with the distance
from the corresponding substrate,
$ \eta(y) = \eta_0 \left[ 1-\tanh (y/y_d) \right]$,
where typically $\eta_0 =1$ and $y_d \sim 1$.

\section{Results}
\label{results}

\begin{figure}[h] 
\begin{center}
\includegraphics[width=8cm]{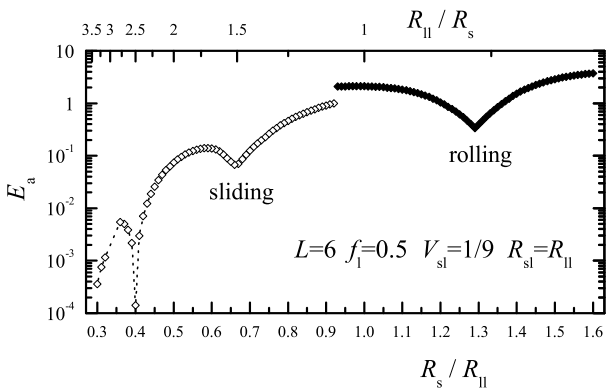} 
\caption{\label{R04c} 
Activation energy $E_a$ as a function of the substrate lattice constant $R_s$
for the rigid $L=6$ molecule, for $f_l=0.5$, $V_{sl}=1/9$, and $R_{sl}=R_{ll}$.
Open symbols correspond to sliding, solid symbols to rolling.}
\end{center}
\end{figure}

\begin{figure}[t] 
\includegraphics[clip,width=7cm]{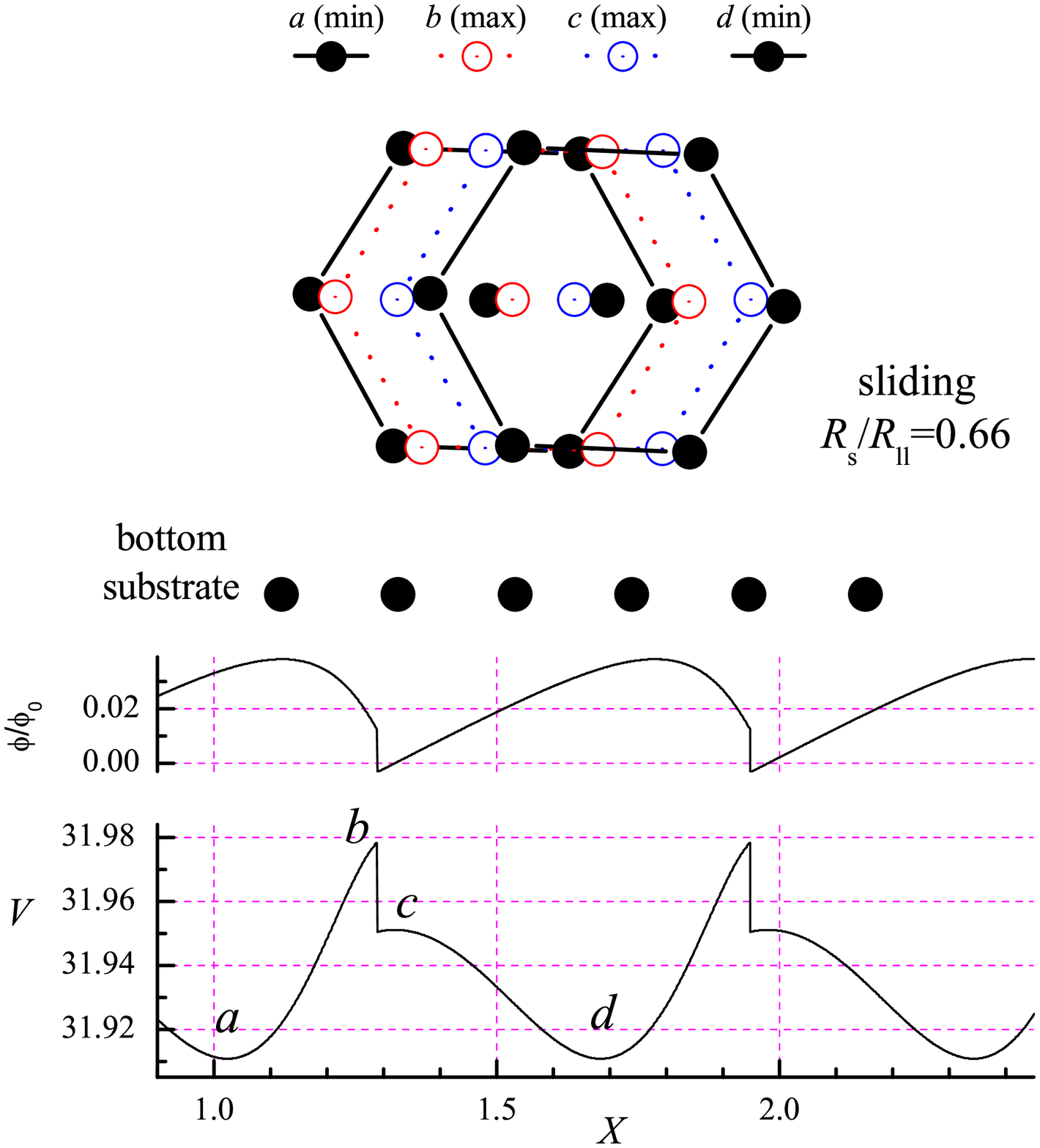} 
\hspace{1cm}
\includegraphics[clip,width=7cm]{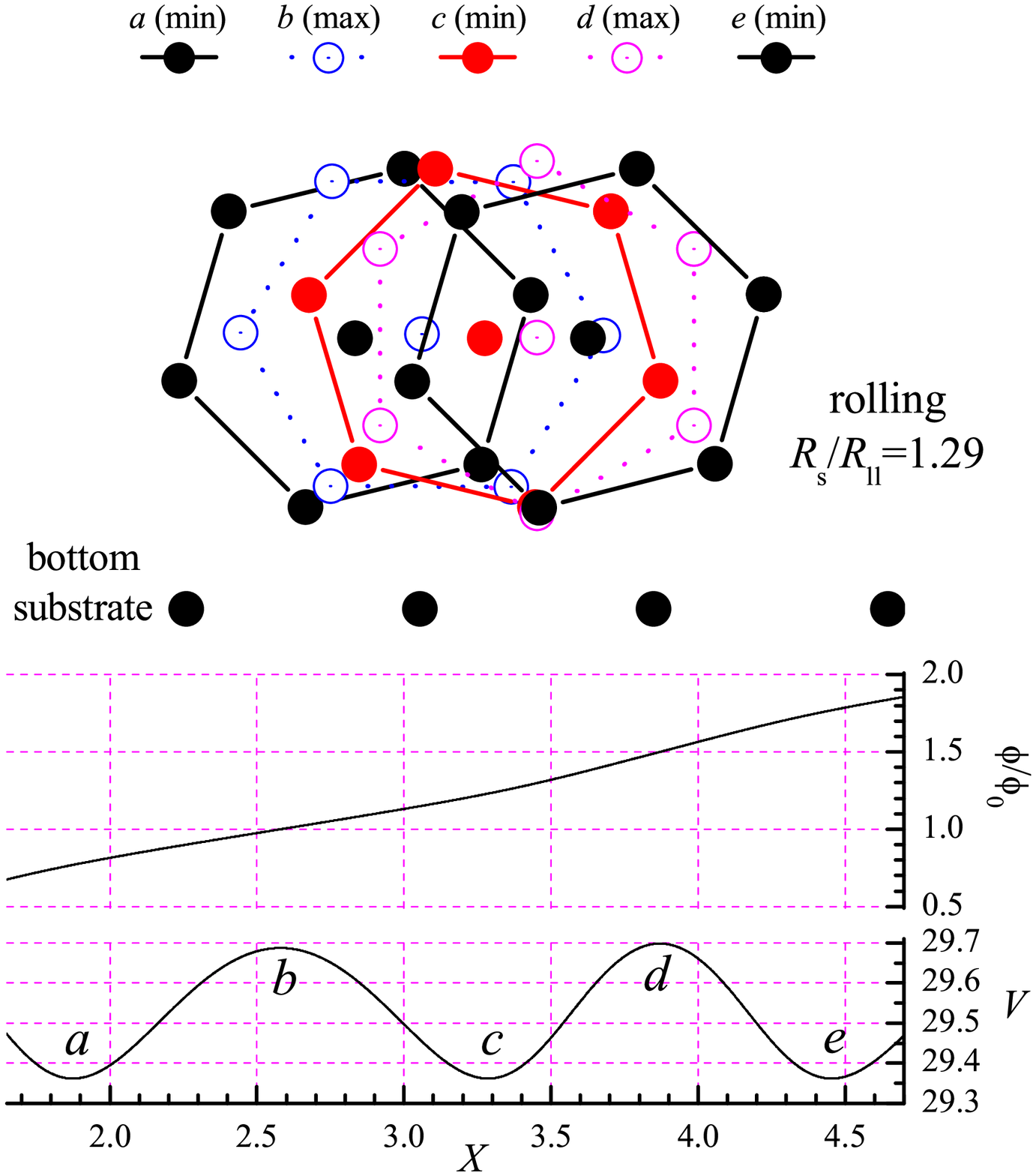} 
\begin{center}
\caption{\label{R05} (color online):
Sliding adiabatic motion of the rigid $L=6$ molecule for $R_s/R_{ll}=0.66$
($\Delta \phi < \phi_0$, left panel)
and rolling for $R_s/R_{ll}=1.29$ ($\Delta \phi > \phi_0$, right panel).
Other parameters as in Fig.~\ref{R04c}.
Lower panels: $X$-dependence of the potential energy $V(X)$,
and the angle $\phi (X)/\phi_0$. Top panel: configurations as the
molecule moves from one minimum of $V(X)$ to the next.}
\end{center}
\end{figure}

\begin{figure}[h]
\includegraphics[width=7.5cm]{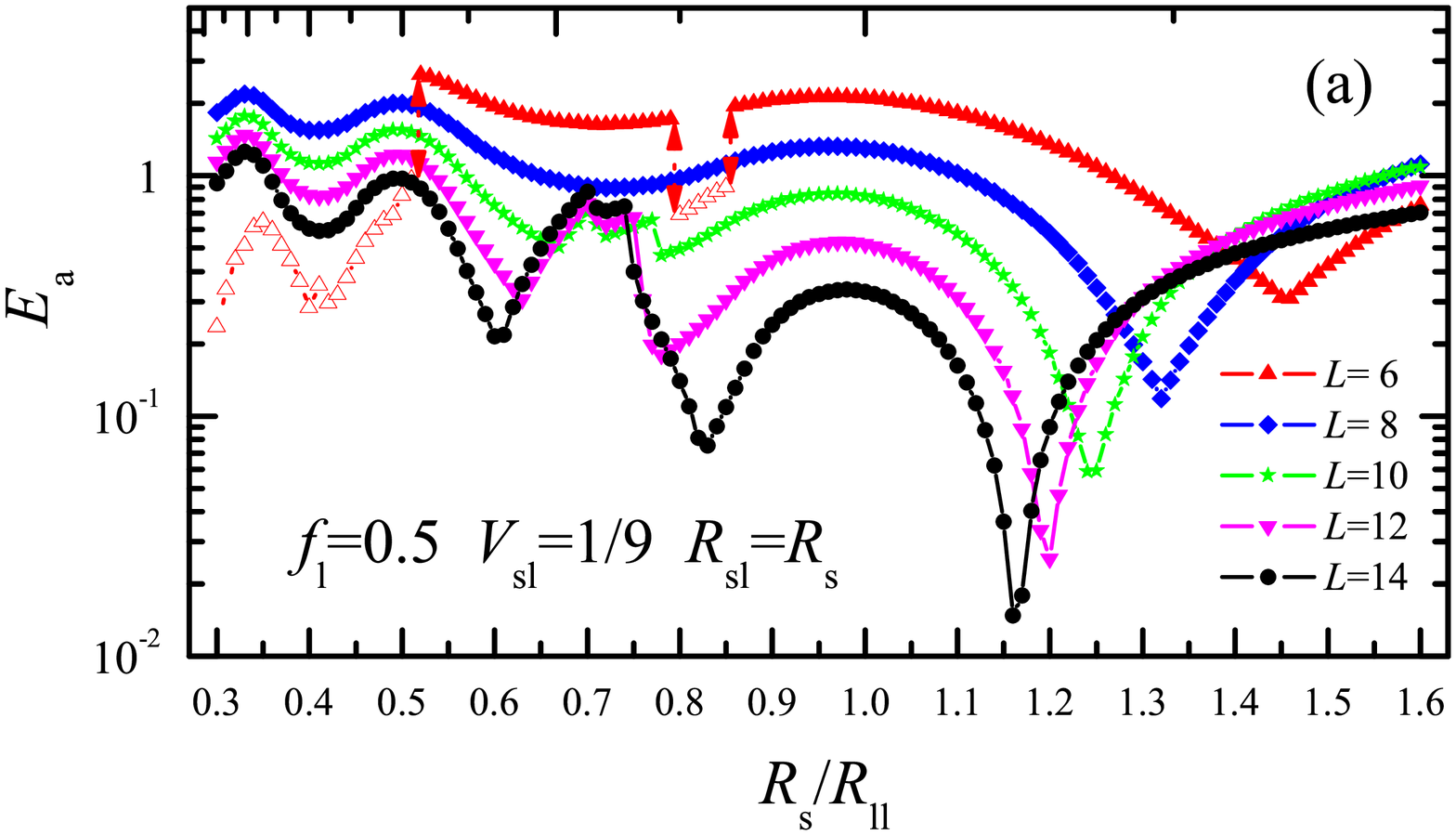}
\includegraphics[width=7.5cm]{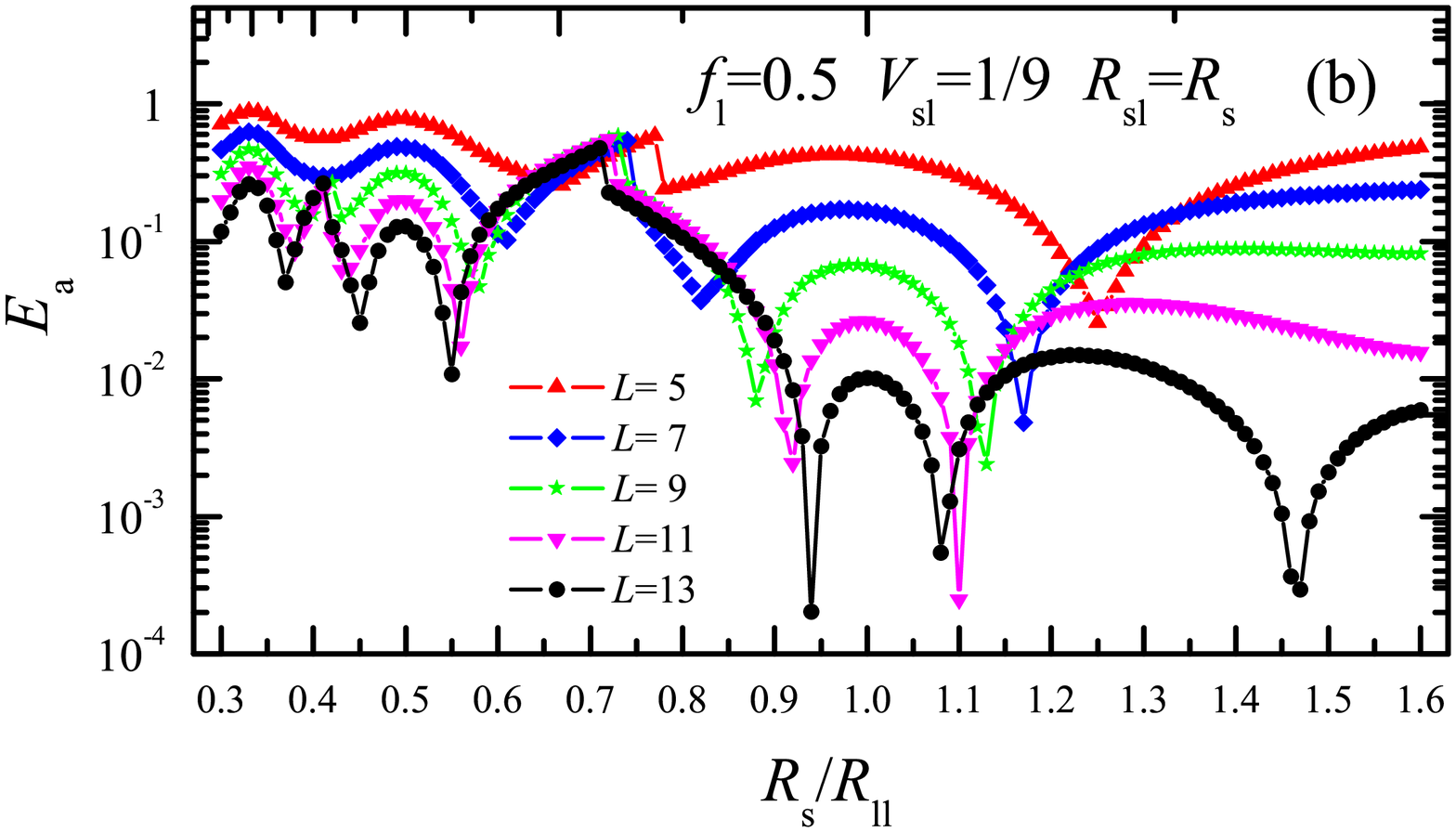} 
\begin{center}
\caption{\label{R07} (color online):
Rigid molecule activation energy $E_a$ versus $R_s /R_{ll}$ for
$f_l =0.5$ and $V_{sl}=1/9$. Unlike Fig.~\ref{R04c}, here $R_{sl}=R_s$.
Panel (a): even $L=6$, 8, 10, 12, 14; panel (b), odd $L=5$, 7, 9, 11, 13.
Empty triangles in $L$=6 indicate sliding friction intervals.}
\end{center}
\end{figure}

\begin{figure}
\begin{center}
\includegraphics[width=7.5cm]{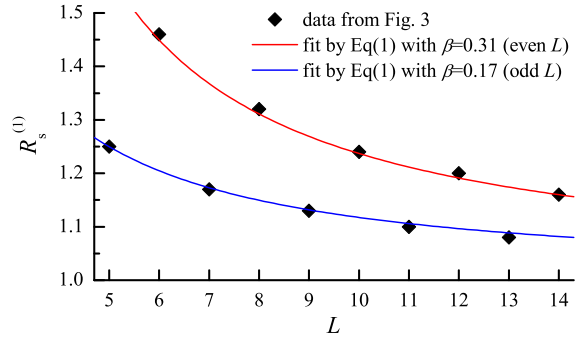} 
\caption{\label{R15} (color online):
Position $R_s^{(1)}$ of the main minimum of $E_a (R_s)$, extracted
from Fig.~\ref{R07}, for increasing molecular size $L$. Curves are fits to
the cogwheel model  Eq.~(\ref{Rsmatch}). The asymptotic limit of 1
is still relatively far for reasonable molecular radii. }
\end{center}
\end{figure}

\begin{figure}
\includegraphics[width=7.5cm]{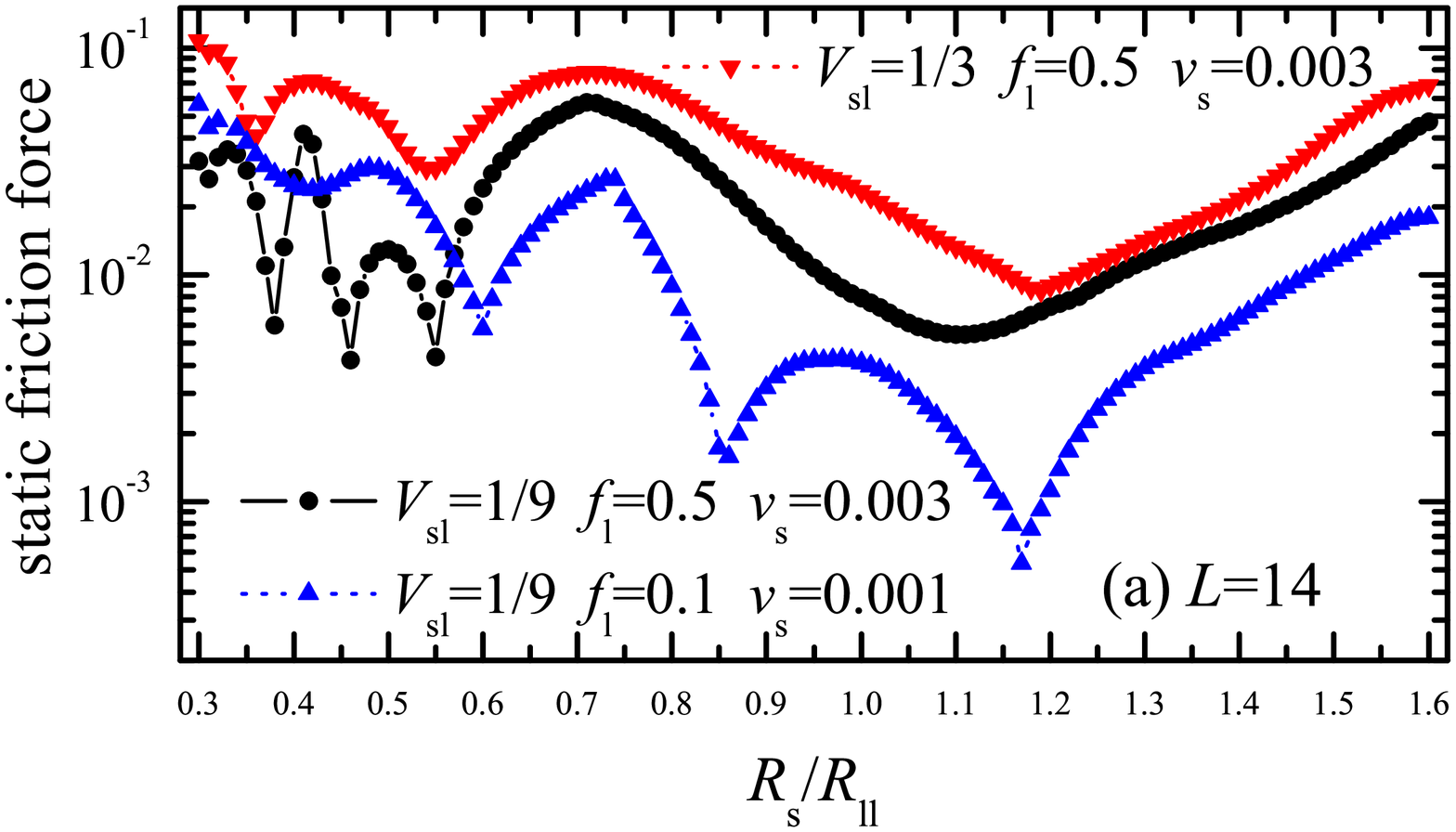} 
\includegraphics[width=7.5cm]{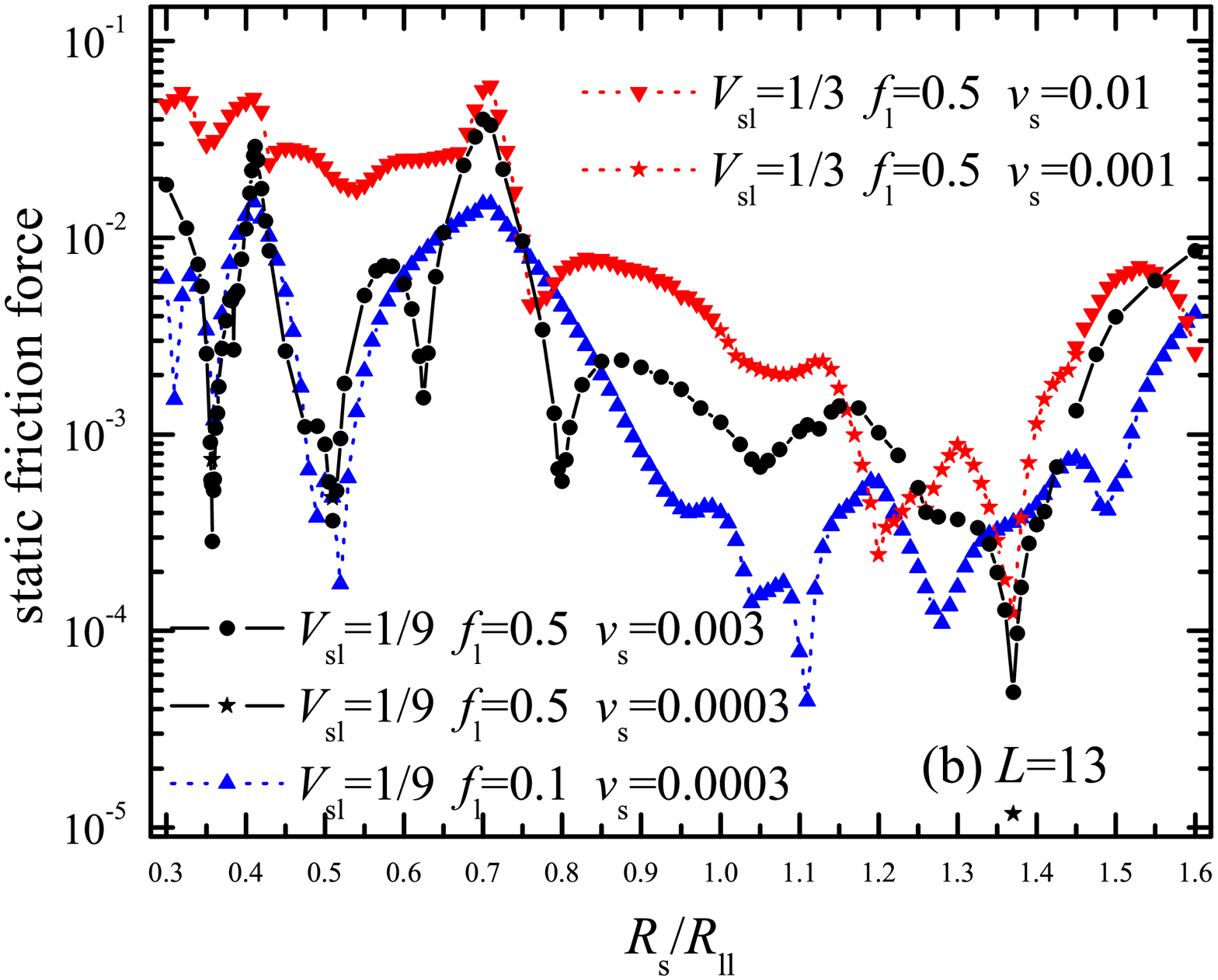} 
\begin{center}
\caption{\label{R09} (color online):
Static friction force $f_s$ as a function of the substrate lattice constant $R_s /R_{ll}$
for (a) $L=14$ and (b) $L=13$ for different system parameters:
({\it i\/})~$f_l =0.5$ and $V_{sl}=1/9$ (solid curve and circles and stars),
({\it ii\/})~$f_l =0.5$ and $V_{sl}=1/3$ (down triangles and red dotted curve and stars), and
({\it iii\/})~$f_l =0.1$ and $V_{sl}=1/9$ (up triangles and blue dotted curve).
Other parameters are: $R_{sl}=R_s$, $K_m=100$ and $V_{ll}=1$.}
\end{center}
\end{figure}

Simulation results for single molecule friction were obtained
from $L=5$ (the simplest circular molecule) up to $L=13$ and~14,
which may be considered as a 2D version of fullerene.
In fact in the 3D case, the surface area of the spherical molecule is $s=4\pi R_m^2$.
With $L_3 =60$ atoms on the surface, this gives $s \approx L_3 R_{ll}^2$,
or $R_m /R_{ll} \approx 2.18$. In 2D, the length of the circle is $2\pi R_m \approx L R_{ll}$,
or $L \approx 2\pi R_m /R_{ll}$, which leads to $L \approx 13.7$
for the same ratio $R_m /R_{ll}$ as for 3D fullerenes.
We first consider a rigid molecule,
$V_{ll} = \infty$ and $K_m = \infty$, fix $X$ of
the top substrate and seek the potential energy minimum $V$ by varying
its $y$-coordinate $Y$, the molecular center $(x_c, y_c)$ and the rotation angle $\phi$.
The $X$ dependence of $V$, $(x_c, y_c)$, and $\phi$ defines the adiabatic
trajectory, which describes the joint substrate and lubricant motion when infinitely slow.
We extract the activation energy $E_a = \max \left[ V(X) \right] - \min \left[ V(X) \right]$,
and the static friction force magnitude, approximated  as
$f_s = \max \left[ dV(X)/dX \right]$ ($f_s \sim E_a$ in our units).

Figures \ref{R04c} and~\ref{R05} show results for the $L=6$ molecule
when, to simplify further, $R_{sl}$ is fixed to $R_{sl}=R_{ll}$.
The energy $V(X)$ is periodic with $R_s$ (or a multiplier of $R_s$).
The molecular angle $\phi$ varies by $\Delta \phi$
as the potential energy $V(X)$ changes from minimum to maximum.
Because $\phi (X)$ must be continuous,
the motion corresponds to sliding if $\Delta \phi < \phi_0 \equiv 2\pi/L$,
while if $\Delta \phi > \phi_0$ the molecule must rotate when it moves.
As Fig.~\ref{R04c} shows,
for $R_s < R_{ll}$ the motion corresponds to sliding, i.e.,
the molecule is shifted as a whole, slightly oscillating during motion
(see Fig.~\ref{R05}, left panel).
The sliding activation energy has maxima at $R_{ll}=nR_{s}$ (where $n$ is an integer)
and minima at $R_{ll}=(n-1/2)R_{s}$.
On the other hand, for $R_{ll} < R_{s}$ the motion corresponds to rolling
(Fig.~\ref{R05}, right panel). Here $E_a(R_s)$ has sharp minima at unanticipated
values of $R_{s}/R_{ll}$ (e.g., for $R_{s}/R_{ll} \approx 1.29$ in Fig.~\ref{R04c}).

Varying $R_s$ in Fig.~\ref{R04c} at fixed value of $R_{sl}$ for the lubricant-substrate
interaction, we find that for $R_s < R_{ll}$ rolling
replaces sliding already around $R_s /R_{ll} =2/3$.
Preference for rolling over sliding increases for increasing load $f_l$
and for decreasing interaction strength $V_{sl}$ \cite{note1}
(choosing alternatively $R_{sl}=R_s$ rolling is even more prevalent
than for fixed $R_{sl}$).
When sliding wins over rolling, 
it generally provides a lower activation energy.
Recalling that $\phi_0=2\pi /L$, the region of parameters
for rolling should increase with $L$ --- a rounder wheel rolls better!
The $R_s$ dependence of $E_a (R_s)$ for increasing size $L$ (Fig.~\ref{R07})
shows rolling for all $R_s$ and for all $L \geq 5$,
except for $L=6$ which shows both rolling and sliding (see open symbols in Fig.~\ref{R07}a).
As $R_s$ varies, the value of $E_a$ oscillates by more than two orders of magnitude
for even $L$ and more than three for odd $L$, with deep sharp minima
separated by broad maxima. Clearly, by suitably choosing
$R_s /R_{ll}$ a very strong decrease of rolling friction is attainable.

The unexpected minima of $E_a (R_s)$ can be explained by simple engineering
--- a ``rack-and-pinion'' model. Consider the molecule as a cogwheel
(the pinion) with $L$ cogs, primitive radius $R_m$
and external radius $R^* = R_m +h$, where $h \propto R_{sl}$.
The chord distance between nearest cogs is $R_{ll}^* = 2R^* \sin (\pi/L)$.
Best rolling conditions are expected when $R_{ll}^*$ matches the
substrate potential period $R_s$, i.e., for $R_s^{(1)} = R_{ll}^*$ and its fractions,
$R_s^{(2)} = R_{ll}^*/2$, $R_s^{(3)} = R_{ll}^*/3$, etc.
The main minimum of $E_a (R_s)$ is expected at
\begin{equation}
R_s^{(1)} / R_{ll} = 1+(2h/R_{ll}) \sin (\pi/L).
\label{Rsmatch}
\end{equation}
As shown in Fig.~\ref{R15}, the cogwheel model (\ref{Rsmatch}) with
$h=\beta R_{sl}$, where 
$\beta$ is a parameter, fits extremely well
the shift of minimum position with molecular size $L$. Moreover, it can
explain its variation with  load (the radius $R^*$ and therefore $h$
decreases as the load grows) as well as with the lubricant-substrate
interaction $V_{sl}$ ($R^*$ and $h$ decreases with $V_{sl}$) \cite{note1}.
It also accounts for the even-odd effect
since odd $L$ involves perfect ingraining for one substrate
at a time,  justifying why roughly double values of $\beta$
are needed for even relative to odd $L$.

These insights for rigid molecules are confirmed
by the static friction force obtained from simulation with deformable molecules
(Fig.~\ref{R09}). The friction coefficient $\mu_s = f_s /f_l$ ranges
from $\mu_s \sim 0.1$ at $R_s /R_{ll} \approx 0.7$
to $\mu_s \sim 0.01$ or even $\mu_s \sim 0.001$ at $R_s /R_{ll} \approx 1.1$.
The results are also robust to changes of model parameters.
For example, Fig.~\ref{R09} compares the dependences $f_s (R_s)$ for two values
of the amplitude of lubricant-substrate interaction,
$V_{sl}=1/9$ and~$1/3$, and for two values of the load,
$f_l =0.5$ and~$0.1$.
We found both static $f_s$ and kinetic friction $f_k$ to increase approximately
linearly with load, $f_{s,k} \approx f_{0s,0k} + \mu_{s,k} f_l$.
Visualization of MD trajectories 
shows that for $R_s /R_{ll}=0.7$, where friction is high,
rolling rotation is in fact discontinuous, and accompanied by a molecular
shift/sliding --- much as cogwheels with excessive clearance would do ---
which dissipates mechanical energy into vibrations, whereas
for $R_s /R_{ll}=1.1$, where friction is low, the motion is a
smooth rotation, corresponding to optimal rack-and-pinion coupling.

Simulations further showed that this scenario remains valid at nonzero temperature $T$.
As $T$ increases, both static and kinetic friction forces are found to decrease,
the stick-slip changing to creep and finally to smooth motion at high temperatures.
We also found transitions from stick-slip to smooth rolling
for increasing velocity. The cogwheel effect remains, and for example
calculated static friction for 
$R_s /R_{ll}=0.7$ and $R_s /R_{ll}=1.1$ still differ by a factor of ten or more.
The critical velocity $v_c$ of the transition from stick-slip to smooth rolling
also differs by a factor of about four in the two cases. In all cases we find
$f_k \ll f_s$ \cite{note1}.

The present approach to the single rolling molecule can be extended
to a finite coverage of lubricant molecules.
For $M$ molecules the individual molecular load is $f_{l1}=f_l /M$ 
so that the friction per molecule $f_{s1,k1}$
also decreases. 
However, the total friction force
$f_{s,k}=M f_{s1,k1}$ increases, so that the combined effect is a slow
increase of the total friction with $M$.
In a real situation, coalescence may lead
to jamming, with molecules blocking their mutual rotation.
In our model, jamming starts already at $\theta_M \approx 0.1$
and completely destroys rolling at $\theta_M > 0.3$~\cite{B2005, note1}
(here $\theta_M =M/M_1$ is the coverage, with $M_1$ the number of
molecules in a monolayer).

\section{Conclusion}
\label{conclusion}

Our 2D modeling leads to overall conclusions of considerable novelty.
Rolling spherical lubricant molecules can indeed provide
better tribological parameters than sliding atomic lubricants.
The effect may be as large as in macroscopic friction,
where rolling reduces friction by a factor of $10^2 - 10^3$,
however only for low concentration of lubricant molecules,
and for specially chosen values of the ratio $R_{s}/R_{ll}$,
e.g., for $R_{s}/R_{ll} \sim 1.1$, corresponding to perfect rack-and-pinion matching.

While of course the matching condition will be more delicate and difficult
to realize in 3D than in 2D, the concept can surely be pursued experimentally,
for different spherical molecules, different substrates, and different coverages.
Inert nonmetal surface (such as perhaps self-assembled monolayers) could represent
a better choice of substrate than metals for fullerene deposition; and lower coverages
should be preferred to 
complete monolayers.
Because 3D rolling has an azimuthal degree of freedom, the novel
cogwheel effect found here will be direction dependent, and rolling
friction will exhibit anisotropy depending on direction.
At fixed rolling direction, increasing load could offer the simplest tool
to change the rack-and-pinion matching through a pressure-induced
decrease of $h=\beta R_{sl}$. In this case, the effect would show up
as a strikingly nonmonotonic (non-Amontons) behavior of friction with load.

\section*{Acknowledgement}
This research was supported in part by MIUR PRIN/Cofin Contract No.~2006022847.
OMB gratefully acknowledges a Central European Initiative (CEI) grant,
and the hospitality of SISSA.


\end{document}